\def\year{2022}\relax
\documentclass[letterpaper]{article} 
\usepackage{aaai22}  
\usepackage{times}  
\usepackage{helvet}  
\usepackage{courier}  
\usepackage[hyphens]{url}  
\usepackage{graphicx} 
\usepackage{natbib}  
\usepackage{caption} 
\DeclareCaptionStyle{ruled}{labelfont=normalfont,labelsep=colon,strut=off} 
\usepackage{algorithm}
\usepackage{algorithmic}

\usepackage{newfloat}
\usepackage{listings}
\floatstyle{ruled}
\newfloat{listing}{tb}{lst}{}
\floatname{listing}{Listing}
\usepackage[utf8]{inputenc} 
\usepackage{booktabs}       
\usepackage{amsfonts}       
\usepackage{nicefrac}       
\usepackage{microtype}      
\usepackage{xcolor}         
\usepackage{tikz}
\usepackage{dblfloatfix}
\usepackage{bbm}
\usepackage{amsmath}
\usepackage{efbox}
\usepackage{makecell}
\usepackage{blindtext}
\usepackage{multirow}
\usepackage[most]{tcolorbox}
\usepackage[percent]{overpic}

\def\argmin{\mathop{\rm argmin}}
\def\argmax{\mathop{\rm argmax}}
\newcommand{\etal}{~\emph{et al.}}
\efboxsetup{linecolor=black,linewidth=2.5pt}

\title{Reference-guided Pseudo-Label Generation for Medical Semantic Segmentation}

\title{Reference-guided Pseudo-Label Generation for Medical Semantic Segmentation}
\author {
    Constantin Seibold,\textsuperscript{\rm 1}
    Simon Rei\ss, \textsuperscript{\rm 1}
    Jens Kleesiek \textsuperscript{\rm 2}
    Rainer Stiefelhagen, \textsuperscript{\rm 1}
}
\affiliations {
    \textsuperscript{\rm 1} Karlsruhe Institute of Technology\\
    \textsuperscript{\rm 2} University Medicine Essen\\
    \textsuperscript{\rm 1}\{firstname.lastname\}@kit.edu, \textsuperscript{\rm 2}\{firstname.lastname\}@uk-essen.de, 
}

\begin{document}

\maketitle

\begin{abstract}
Producing densely annotated data is a difficult and tedious task for medical imaging applications.
To address this problem, we propose a novel approach to generate supervision for semi-supervised semantic segmentation. We argue that visually similar regions between labeled and unlabeled images likely contain the same semantics and therefore should share their label. Following this thought, we use a small number of labeled images as reference material and match pixels in an unlabeled image to the semantics of the best fitting pixel in a reference set.
This way, we avoid pitfalls such as confirmation bias, common in purely prediction-based pseudo-labeling.
Since our method does not require any architectural changes or accompanying networks, one can easily insert it into existing frameworks.
We achieve the same performance as a standard fully supervised model on X-ray anatomy segmentation, albeit \emph{95\% fewer labeled images}.
Aside from an in-depth analysis of different aspects of our proposed method, we further demonstrate the effectiveness of our reference-guided learning paradigm by comparing our approach against existing methods for retinal fluid segmentation with competitive performance as we improve upon recent work by up to 15\% mean IoU.
\end{abstract}

\section{Introduction}
\begin{figure}[t]
    \centering
    \includegraphics[width=0.85\linewidth, height=0.85\linewidth]{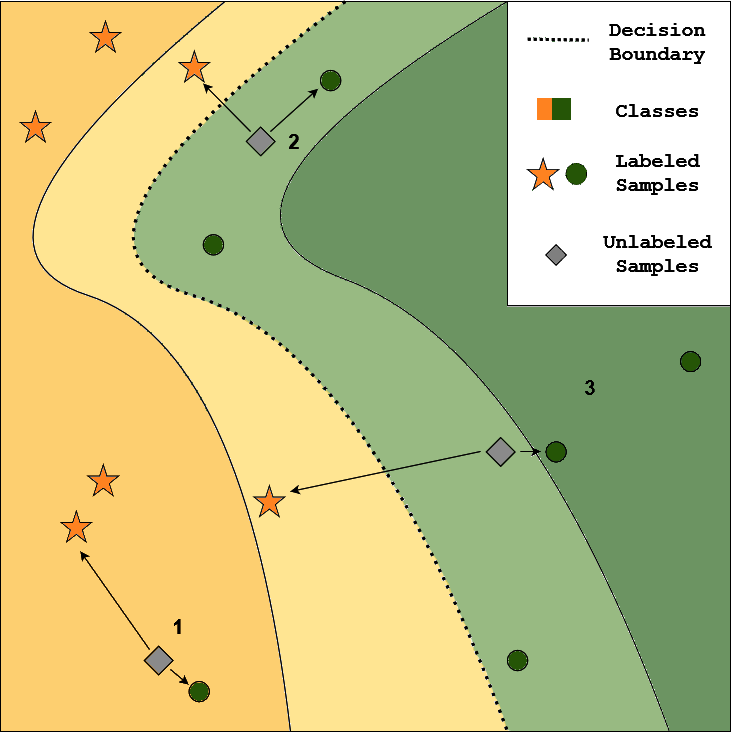}
    \caption{A conceptual example highlighting different cases how unannotated samples could be integrated into a networks feature spaces and decision boundary. Color intensity represents network prediction confidence.}
    \label{fig:intro}
\end{figure}

The acquisition of detailed annotations for semantic segmentation is a complex and time-consuming process~\cite{cordts2016cityscapes}.
It becomes increasingly difficult in the medical domain due to the required expertise~\cite{menze2014multimodal}. 
When considering a doctor's obligations in the clinical routine, gathering a large amount of detailed medical annotations can become almost insurmountable.
Thus, these obstacles make it desirable to perform accurate 
semantic segmentation while minimizing the necessary annotated data.

Semi-supervised semantic segmentation solves these tasks by combining small quantities of labeled data with a lot of unannotated data for training.
In recent years, several directions have been investigated such as student-teacher frameworks~\cite{gou2021knowledge,chen2020big,Xie_2020_CVPR,pham2021meta}, consistency regularization~\cite{ouali2020semi,sohn2020fixmatch,rebuffi2020semi} or pseudo-labels ~\cite{lee2013pseudo,iscen2019label,rizve2021defense}. 
Most pseudo-label methods typically employ network predictions for unlabeled data to either save them for retraining or use them online as targets in the same iteration. They are often paired with the generation of predictions from different perturbations, e.g. data-augmentations, on an input image~\cite{berthelot2019mixmatch,berthelot2019remixmatch,sohn2020fixmatch}.

However, by enforcing predictions' pre-existing biases of the network, incorrect conclusions can have a snowballing negative effect. We run into the issue of confirmation bias~\cite{rizve2021defense}.
Yet, we argue that the positive properties of these methods can be kept, while reducing the adverse effects by taking a different path: comparing embeddings between predictions of unlabeled samples and labeled reference images to instill the supervision.
For all pixels in a given unlabeled image, we find class-wise nearest-neighbors among the pixels of images in the small labeled reference set.
From this, we compute a class proximities, attribute importance via confidence-based weighting and then infer the pseudo-label.
By taking this detour and not directly using the class-predictions, but matching them to a known reference set, the bias towards large classes can be regulated and  we bypass the problems of direct class predictions as supervision.

We illustrate the characteristics in Fig. \ref{fig:intro}. 
First, while the \textit{U}nlabeled \textit{P}ixel (UP), depicted by grey-diamond 1, would be predicted as orange, it is more similar to the misclassified green sample, and due to its proximity, we would instead pass a green pseudo-label.
In the second case, we would assign a green pseudo-label to the UP2, but as its distance to both an orange and a green sample is nearly equal, the weighting would be minimal.
In the third case, we assign a green pseudo-label with a high weight to the UP as it is close to a green sample and far from the second nearest class.

We view this approach as a straightforward support mechanism to pseudo-labeling in semi-supervised semantic segmentation.
It is easily extended with any semi-supervised learning approaches.
We demonstrate the effectiveness of our methods with extensive experiments for multi-class and binary multi-class semi-supervised semantic segmentation on the RETOUCH~\cite{retouch} and JSRT~\cite{jsrt} datasets.
We achieve competitive results across these datasets, excelling especially for minimal amounts of samples. 
We summarize our contributions as:
\begin{enumerate}
    \item We illustrate a different view on online pseudo-labels in semantic segmentation. By enforcing consistency between predictions and the feature space, we cover cases not handled by standard pseudo-labeling approaches. 
    \item We show the effectiveness of our method on different datasets and various low data settings. Thereby, we demonstrate its use for handling the challenging segmentation of overlapping labels from scarce data as we reach fully supervised performance from six labeled images.
    \item We provide a detailed ablation study investigating different aspects of our pseudo-labels in various settings. 
\end{enumerate}

\section{Related Work}
\noindent\textbf{Semi-Supervised Learning.} Semi-supervised learning (SSL) utilizes few labeled samples paired with unlabeled samples to perform a given task. 
Recently, this field has seen significant progress~\cite{berthelot2019mixmatch,berthelot2019remixmatch,cascante2020curriculum,ouali2020semi,chen2020big,sohn2020fixmatch,pham2021meta}. Most methods follow one or combinations of directions such as entropy minimization~\cite{grandvalet2005semi}, consistency regularization~\cite{tarvainen2017mean,ji2019invariant,sohn2020fixmatch} or pseudo-labeling~\cite{lee2013pseudo,iscen2019label,cascante2020curriculum}. Pseudo-labeling-based approaches typically train a classifier with unlabeled data using pseudo targets derived from the model’s high-confidence predictions\cite{lee2013pseudo}. However, pseudo-labeling can lead to noisy training due to poor calibration and as a result of incorrect high-confidence predictions~\cite{guo2017calibration,rizve2021defense}. Other methods approach pseudo-labeling by following a transductive setting, i.e.~setting up a nearest-neighbor graph and perform label propagation. This generation process of pseudo-labels is not feasible in an online setting due to the high demand on run-time and memory consumption for label-propagation and is performed after a set amount of iterations  ~\cite{shi2018transductive,iscen2019label,liu2019deep}. 
In this fashion, pseudo-labeling literature can be divided into online variants, which build pseudo-labels for unlabeled data directly during forward pass~\cite{lee2013pseudo,sohn2020fixmatch}, and offline variants, which generate new targets for the dataset in greater intervals~\cite{iscen2019label,chen2020big, xie2020self,cascante2020curriculum, pham2021meta}. Recently, \cite{taherkhani2021self} matches clusters of unlabeled data to their most similar classes in an offline procedure.
Taking the advantageous aspects of label-propagation methods~\cite{shi2018transductive,iscen2019label,liu2019deep}, we introduce a way to make them work online and even for semantic segmentation where \emph{pixel-wise} labels add to the computational load.
Favorable storage requirements of our solution make a streamlined integration with consistency regularization methods possible.
In consistency regularization, predictions for varied versions of the same input are enforced to be similar.
Usually this is achieved by setting up augmented versions of an input image~\cite{sohn2020fixmatch}, perturbations of feature maps~\cite{ouali2020semi} or different network states~\cite{tarvainen2017mean}.
In our work, we intertwine online-generated pseudo-labels with consistency regularization to alleviate drawbacks in either of the two.

\noindent\textbf{Semi-Supervised Segmentation.}
Semi-supervised semantic segmentation proposed several extensions to concepts in SSL, i.e., to consistency regularization. \cite{french2019consistency} integrate CutMix~\cite{zhang2017mixup} to enforce consistency between mixed outputs and the prediction from corresponding mixed inputs. CCT~\cite{ouali2020semi} aligns the outputs of the main segmentation decoder module and auxiliary decoders trained on different perturbations to enforce consistent feature representations. PseudoSeg~\cite{zou2021pseudoseg} adapts FixMatch~\cite{sohn2020fixmatch} and thus enforces consistency between segmentations of weakly and strongly augmented images employing GradCAM~\cite{selvaraju2017grad}. \cite{chen2021semi} use two independent networks with the same structure and enforce consistency between  their predictions.
In contrast to these approaches, our method utilizes labeled data in pseudo-label generation without any network alterations, rendering it flexible to integrate.

\noindent\textbf{Self-Training for Medical Imaging.}
Network-produced supervision for unlabeled data is starting to gain traction in medical image analysis. \cite{tang2021leveraging} propose a self-training-based framework for mass detection in mammography leveraging medical reports. \cite{NEURIPS2020_Chaitanya} depict the use of contrastive self-supervised learning for semi-supervised segmentation. \cite{Seibold_2020_ACCV} devise entropy minimization to localize diseases in chest radiographs while \cite{huo2021atso} propose a student-teacher framework for semi-supervised pancreas tumor segmentation. ~\cite{reiss2021every} propose to use deep supervision, which adapts networks to enforce prediction consistency between different layers of the network. However, most methods either utilize network predictions or predefined signals~\cite{ouyang2020self} as supervision.
Our method is suitable for 
medical image analysis due to inherent structural similarities, e.g. anatomical properties make a natural fit for our method as labeled images can convey strong reference points for matching and transferring information onto unlabeled images.

\begin{figure*}[t!]
    \centering
        \begin{tabular}{c}
            \includegraphics[width=\linewidth]{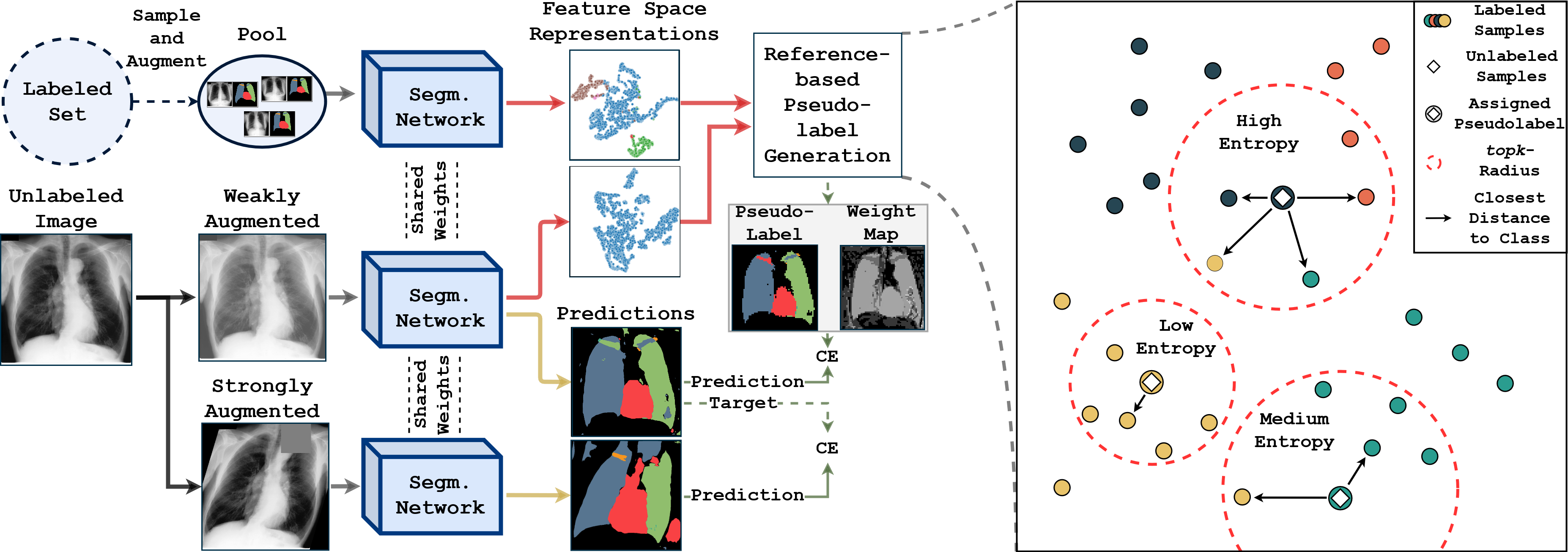}
        \end{tabular}
    \caption{Overview of the proposed training step for unlabeled images. For each unlabeled image, we extract its features in addition to a pool of sampled annotated images to generate pseudo-labels. In parallel, we use the predictions of a weakly augmented sample to act as supervision for a strongly augmeted version of the image. On the right, we illustrate our reference-based pseudo-label generation process for $k=5$. }
    \label{fig:overview}
\end{figure*}
\section{Methodology}
In this section, we propose a novel strategy to generate online pseudo-labels based on label-wise feature similarities from a pool of references. We first define preliminary information, then elaborate on Reference-based Pseudo-Label Generation (RPG) and finally expand on RPG with augmentation-based consistency regularization.
\subsection{Preliminaries}
In the setting of semi-supervised semantic segmentation, a small set of labeled $\mathcal{S_L} = \{(x_i,y_i)\}^{N_l}_{i=1}$ and a large amount of unlabeled images $\mathcal{S_U} = \{x_i\}^{N_u}_{i=1}$ are provided. An image be defined as $x_i \in \mathbb{R}^{ch\times h\times w}$ with $ch$ image channels, height $h$ and width $w$.
Labels be defined as $y_i\in \{0,\dots,c-1\}^{h\times w}$ in case of segmentation into $c$ classes or $y_i\in \{0,1\}^{c\times h\times w}$ if at each location more than one class can be present (multi-label segmentation).
Thus, the task resolves to using $\mathcal{S_L}$ and $\mathcal{S_U}$ to find a model that correctly predicts labels on unseen images.
For later purposes, we define the segmentation model as (1) a dense feature extractor $f_{\text{feat}}:\mathbb{R}^{ch\times h\times w} \rightarrow \mathbb{R}^{d\times h\times w}$ and (2) a subsequent pixel-wise classifier $f_{{cls}}:\mathbb{R}^{d \times h\times w} \rightarrow [0,1]^{c\times h\times w}$ that transforms the $d$ dimensional features at each location into class predictions.
$f_{{\text{feat}}}$ is parameterized by a neural network and for $f_{{\text{cls}}}$ we leverage a $1\times1$ convolution and normalization function (sigmoid or softmax depending on the $y_i$ formulation).

\subsection{Reference-based Pseudolabel Generation}
We propose using image-reference pairs in the pseudo-label generation process for semantic segmentation.
Contrary to directly deriving pseudo-labels from network predictions, we search for a best fit in feature space among a pool of labeled reference images and transfer their semantics.
We display our approach in Fig.~\ref{fig:overview}.

\noindent\textbf{Reference Pool.}
We utilize labeled references to generate pseudo-labels. Therefore, we project both labeled and un-labeled pixels into the same feature space using $f_{\text{feat}}$.
Since available memory is limited, processing $h \times w$ $d$-dimensional pixel-wise representations for each image in $\mathcal{S_L}$ is unfeasible. Additionally, solutions like a memory-bank~\cite{wu2018unsupervised} are difficult to integrate due to sheer amount of pixel-wise representations.
We approach these issues by randomly sampling a pool $\mathcal{P}$ of labeled images from $\mathcal{S_L}$ in each mini-batch iteration:
\begin{equation}
 \mathcal{P}=\{(x,y) \sim \mathcal{S_L}\}^p 
\end{equation}
As we later generate pseudo-labels from $\mathcal{P}$, all classes have to be present otherwise the missing class-labels can not be recovered.
We, thus, sample $p$ images such that each class occurs at least once in $\mathcal{P}$.

We generate our reference set $\mathcal{R}_{\mathcal{P}}$ by extracting the pixel-wise features of each image in $ \mathcal{P}$ to get pairs of pixel-representations and -labels:
\begin{equation}\mathcal{R}_{\mathcal{P}}=\{(f_\text{feat}(x),y): (x,y) \in \mathcal{P}\}.\end{equation} 
To further reduce the memory constraints we sub-sample the pixel-wise representations and labels to a feasible size $s \times s$ using nearest-neighbor interpolation.
In the following, we dispose of the spatial relations between pixels and only consider $\mathcal{R}_{\mathcal{P}}$ to be a set of $d$ dimensional feature vector-label pairs with $|\mathcal{R}_{\mathcal{P}}| = p\cdot h\cdot w$.
 By sampling $\mathcal{P}$ continuously during training, the labeled images can experience a large variation of data augmentation techniques, leading to more diverse pixel-representations in the reference set $\mathcal{R}_{\mathcal{P}}$.



\noindent\textbf{Label Assignment.}
We build pseudo-labels by finding the closest labeled pixels in feature space from the reference pool $\mathcal{R}_{\mathcal{P}}$ for each unlabeled pixel.
For each unlabeled image $u \in \mathcal{S}_\mathcal{U}$ in the mini-batch we extract pixel-wise features $\hat u = f_{\text{feat}}(u)$.
We now assign the target of an unlabeled vector $\hat u_{\text{x},\text{y}} $ with the spatial coordinates $\text{x},\text{y} \in \mathbb{N}^{h \times w}$ based on the contextually closest feature vector in $\mathcal{R}_{\mathcal{P}}$.
The clipped cosine distance $\mathcal{D}$ between  of the labeled pixel-representations $r\in \mathcal{R}_{\mathcal{P}}$ and the unlabeled pixel-representations $\hat u_{\text{x},\text{y}} \in \hat u$ serves as proximity measure:
\begin{equation}
    \mathcal{D}(r,\hat u_{\text{x},\text{y}})  = 1 - \max(\frac{\sum_{i=1}^{d} r_i\cdot \hat u_{\text{x},\text{y},i}}{\sqrt{\sum_{i=1}^{d} r_i^2}\cdot \sqrt{\sum_{i=1}^{d}\hat u_{\text{x},\text{y},i}^2} + \epsilon },0),
\end{equation}
with subscript $i$ indexing the $i$-th dimension of a vector and the small constant $\epsilon=1e^{-8}$.
Using $\mathcal{D}$, two feature vectors have a distance of zero if they are identical and the maximum distance of one if orthogonal or contrary to each other.
For each unlabeled pixel $u_{\text{x},\text{y}}$ a pseudo-label $l(u_{\text{x},\text{y}})$ is assigned based on the label of its closest sample in the reference pool:
\begin{equation}\label{eq:pixellabel}
    l(u_{\text{x},\text{y}}) = y: \argmin_{(r,y) \in \mathcal{R}_\mathcal{P}} \mathcal{D}(r,\hat u_{\text{x},\text{y}})
\end{equation}
The whole image is labeled by $l(u) = \{l(u_{\text{x},\text{y}}): u_{\text{x},\text{y}} \in u\}$.
Note that this way, $y$ can be either a one-hot vector or a sophisticated multi-label vector.
Our approach is contrary to classical pseudo-labeling, where assigning a multi-label vector requires the network to hit manually designed thresholds for every class.
Our nearest-neighbor target assignment is related to previous methods~\cite{iscen2019label,liu2019deep, mechrez2018contextual}, however, we operate online and access multiple reference images at the same time.


\noindent\textbf{Density-based Class Entropy.}
Overall, for an adequate pool-size $p$, this nearest-neighbor label assignments showed to be beneficial for semantic segmentation.
However, we noticed a potential pitfall: For features with similar distances to several classes direct assignments mislead the network during training.
To avoid this issue, we apply a weighting mechanism based on the ambiguity of an unlabeled pixel's surroundings.
With the feature $\hat u_{\text{x},\text{y}}$ we compute the closest distances $\delta^j_{\hat u_{\text{x},\text{y}}}, j\in {1,\dots,c}$ to each class among the $k$ nearest neighbors $\mathcal{R}_\mathcal{\mathcal{P}}^k$ in feature space. 
\begin{equation}\label{eq:dist_entr}
    \delta^j_{\hat u_{\text{x},\text{y}}} = 
        \min_{(r,y) \in \mathcal{R}_\mathcal{P}^k\land y=j} \mathcal{D}(r, \hat u_{\text{x},\text{y}})
\end{equation}
If class $j$ is not represented in the reference pool $\mathcal{R}_\mathcal{P}^k$, it's distance $\delta^j_{\hat u_{\text{x},\text{y}}}$ is set to one.
We use these class distances to model $j$ class probabilities $\mathit{P}^j_{u_{\text{x},\text{y}}}$ via class-wise normalization:
\begin{equation}\label{eq:class_prob}
    \mathit{P}^j_{u_{\text{x},\text{y}}} = \frac{1- \delta^j_{u_{\text{x},\text{y}}}+\epsilon}{\sum_{j'=1}^c 1- \delta^{j'}_{u_{\text{x},\text{y}}}+\epsilon} 
\end{equation}
We then calculate the weighting factor $W_{u_{\text{x},\text{y}}}$ through the normalized entropy of the class probabilities:
\begin{equation}\label{eq:entropy}
    W_{u_{\text{x},\text{y}}} = 1 + \sum^c_{j=1} \mathit{P}^j_{u_{\text{x},\text{y}}} \frac{\log \mathit{P}^j_{u_{\text{x},\text{y}}}}{\log c}
\end{equation}
With the factor $W_{u_{\text{x},\text{y}}}$ we put a lower weight on pseudo-labeled pixels that lie in highly ambiguous regions in the feature space.
On top this weighting nudges the pseudo-labels towards including more classes instead of opting just for the most common one.
This is due to the fact, that distances of all classes influence the weighting of a pixel-label-assignment.
Further, this weighting handles extreme cases where e.g.  $\delta^j_{u_{\text{x},\text{y}}} = 1$ for all classes, as here the entropy will be maximal, which in turn will lead to ignoring $u_{\text{x},\text{y}}$ since $W_{u_{\text{x},\text{y}}} = 0$.
We illustrate further cases on the righthand side of Fig~\ref{fig:overview}.

Ultimately, our method is formulated as the following loss function $\mathcal{L}_{RPG}$:
\begin{align}
    \mathcal{L}_{RPG} = &
    \, \mathbb{E}_{(x,y)\in \mathcal{S}_\mathcal{L}} 
    [\mathrm{CE}(f_{\text{cls}}^c(f_{feat}(x)),y) ]
    \nonumber\\ \quad
    & +
    \mathbb{E}_{x\in \mathcal{S}_\mathcal{U}} 
    [\mathrm{CE}(f_{\text{cls}}^c(f_{feat}(x)),l(x)) \cdot W_{x}],
\end{align}
with $\mathrm{CE}$ denoting binary or multi-class cross-entropy depending on the type of segmentation task.

\subsection{Consistency Regularization}
To showcase that our approach works complementary to consistency regularization methods in semantic segmentation, we expand the formulation of FixMatch~\cite{sohn2020fixmatch}.
We generate pseudo-labels from network predictions on weakly augmented images and use them as labels for strongly augmented versions of the same image, thereby enforcing consistency between them. 
While weak augmentations are commonly used perturbations such as random flipping, for strong augmentations, we follow RandAugment~\cite{cubuk2020randaugment}.
We follow a similar setting as in Sohn~\etal~\cite{sohn2020fixmatch}.
Since we handle segmentation instead of classification which is done in the original work, we generate pixel-level pseudo-labels and set the designated label for the areas affected by the CutOut augmentation~\cite{devries2017improved} to `background'. 
For one-hot targets $y$, we use the standard pseudo-label formulation~\cite{sohn2020fixmatch}
\begin{equation}\label{eq:multi-class}
    l'( u_{\text{x},\text{y}}) = 
    \begin{cases}
        \argmax_{c} f_{cls}^c(\hat u_{\text{x},\text{y}})       &         \text{, if } f_{cls}^c(\hat  u_{\text{x,y}})>\tau
    \\
    \text{ignore} & \text{, else}
    \end{cases}
\end{equation}
and further extend the FixMatch formulation to enable multi-label segmentation as follows:
\begin{equation}\label{eq:multi-label}
    l'( u_{\text{x},\text{y}}) = 
    \begin{cases}
        \lfloor f_{cls}^c(\hat u_{\text{x},\text{y}}) \rceil 
         & \text{, if } |f_{cls}^c(\hat  u_{\text{x},\text{y}}) - 0.5 | \\ & \quad \;\; > |0.5 - \tau|
    \\
    \text{ignore} ,& \text{, else}
    \end{cases}
\end{equation}
where $\tau$ is a scalar threshold value separating labeled and ignored pixels.
The whole image is labeled by choosing based on the task the respective $l'(\cdot)$ 
$l'(u) = \{l'(u_{\text{x},\text{y}}) : u_{\text{x},\text{y}} \in u\}$.
We denote the final consistency regularized loss term $\mathcal{L}_{{RPG}^+}$ as:
\begin{equation}
\begin{aligned}
    \mathcal{L}_{{RPG}^+} = & \mathcal{L}_{RPG} \,\,+\\&
    \mathbb{E}_{x\in \mathcal{S}_\mathcal{U}}
    [\mathrm{CE}(f_{cls}^c(f_{feat}(\mathit{a}_s(x))),\mathit{a}_s(l'(x))) ]
\end{aligned}
\end{equation}

\begin{table*}[bp]
  \centering
  \begin{tabular}{lllll}
  \toprule
  Methods&\multicolumn{1}{c}{$N_l=3$}&\multicolumn{1}{c}{$N_l=6$}&\multicolumn{1}{c}{$N_l=12$}&\multicolumn{1}{c}{$N_l=24$}\\
    \midrule
    Baseline &$0.59\pm0.04$&$0.73\pm0.02$&$0.81\pm0.01$&$0.85\pm0.01$\\
    Pseudolabel$_{\tau = 0.8}$~\cite{lee2013pseudo}  &$0.56\pm0.04$&$0.73\pm0.04$&$0.81\pm0.03$&\underline{$0.87\pm0.01$}\\
    Pseudolabel$_{\tau = 0.95}$~\cite{lee2013pseudo} &$0.57\pm0.03$&$0.74\pm0.03$&$0.82\pm0.02$&\underline{$0.87\pm0.01$}\\
    Nearest Neighbor                        &$0.64\pm0.05$&$0.76\pm0.02$&$0.81\pm0.02$&$0.84\pm0.01$\\
    FixMatch$_{\tau = 0.8}$~\cite{sohn2020fixmatch}  &\underline{$0.71\pm0.05$}&\underline{$0.79\pm0.02$}&$0.80\pm0.01$&$0.85\pm0.00^*$\\
    FixMatch$_{\tau = 0.95}$~\cite{sohn2020fixmatch} &$0.67\pm0.05$&$0.77\pm0.02$&$0.81\pm0.02$&$0.85\pm0.01^*$\\
    RPG    (Ours)                                 &\underline{$0.71\pm0.02$}&\underline{$0.79\pm0.02$}&\underline{$0.83\pm0.02$}&\underline{$0.87\pm0.01$}\\
    RPG$^+$ (Ours)& $\mathbf{0.77\pm0.05^*}$&$\mathbf{0.85\pm0.01^*}$&$\mathbf{0.87\pm0.00^*}$& $\mathbf{0.88\pm0.01^*}$\\\midrule
    Full Access ($N_l=123$) &\multicolumn{4}{c}{0.85}\\\bottomrule
    
  \end{tabular}
  
  \caption{Performance comparison of our work to related work on the datasets JSRT. $^*$ denotes that due to lacking convergence the model was trained twice the iterations. Bold and underlines denote best and second best performance.}
  \label{table:jsrt_results}

\end{table*}

\section{Experiments}
\subsection{Experimental Setup}
\noindent\textbf{Datasets.}  We evaluate our method on two medical tasks, namely chest radiograph anatomy segmentation and retinal fluid segmentation. For multi-label anatomy segmentation, we employ the public JSRT-dataset~\cite{jsrt}. It consists of five potentially pixel-wise overlapping classes (\textit{right/left clavicle, right/left lung, heart}). The dataset officially exists with two sets of images (123/124). For each amount of labeled images, we choose to generate five distinct random splits from the first set using $N_l$ labeled images ($N_l\in{3,6,12,24}$). 
For each split, we use five images of the first set for validation while using the second set for testing.

To further display the effect of our proposed method for overlapping labels in small data setting, we expand upon the JSRT dataset by using more fine-grained annotations for a total of 72 labels belonging to the supercategories of \textit{heart}, \textit{lung}, \textit{ribs}, \textit{spine}, and \textit{others}. As a result, each label consists of fewer annotated pixels than the large labels in JSRT, i.e., the lung consists of its five lobes compared to two lung halves.
For this task, a medical expert annotated two chest radiographs taking up to 3 hours per image. 
We evaluate the performance on this task by fusing our fine-grained classes into the corresponding JSRT labels , e.g. \textit{right upper}, \textit{middle} and \textit{lower lung lobe} correspond to the \textit{right lung}. We use five JSRT labeled images of the first set as validation and test the performance on the second set of JSRT. We further elaborate on the annotations in the supplementary.

For multi-class retinal fluid segmentation, we utilize the Spectralis vendor of the RETOUCH data set consisting of 14 optical coherence tomography volumes with 49 b-scans each. We follow the setup of \cite{reiss2021every} and thus perform 10-fold cross-validation with training sets using $N_l$ labeled images ($N_l\in{3,6,12,24}$), with validation and test sets of roughly equal size on Spectralis, and ensure that in each split, all diseases show at least once in the mask labels.

 We use mean Intersection over Union (mIoU) as the performance metric. We evaluate our method every tenth epoch, apply the best-performing model on the validation set to the test set, and report the mean and standard deviation.

\noindent\textbf{Implementation Details.}
For our segmentation model we use the common UNet architecture~\cite{ronneberger2015u} with batchnorm~\cite{ioffe2015batch} and bilinear up-scaling blocks. The function $f_{\text{feat}}$ which is used for feature extraction describes the network up to the penultimate layer. The function $f_{\text{cls}}$ is a $1\times1$ convolution followed by a sigmoid function for JSRT and Softmax for RETOUCH. We initialize the network using standard Xavier initialization~\cite{glorot2010understanding}. We optimize using Adam~\cite{kingma2014adam} with learning rate and weight decay of $0.0005$ for 100 epochs on JSRT, 200 on extended JSRT and 50 epochs on RETOUCH respectively. As data augmentations, we use random cropping, rotation, additive noise, and color jitters with additional random flipping. For JSRT, we use batch size $5$ with image size 512, while for RETOUCH we use batch size $8$ following the preprocessing utilized in \cite{reiss2021every}. For all experiments, we build each batch as a combination of $\mathcal{P}$ with $p=3$ and randomly sampled images of the whole dataset. We set the number of considered nearest neighbors $k=7000$ and the representation map size $s=64$. All experiments were run on one 11GB NVIDIA GeForce RTX 2080.

\begin{table}[tp]
  
  \centering
  \footnotesize
  \begin{tabular}{ccccc}
    \toprule
      $p$ &$N_l=3$ & $N_l=6$ & $N_l=12$ & $N_l=24$ \\ \midrule
     1 & $0.52\pm0.02$ & $0.66\pm0.04$ & $0.74\pm0.02$ & $0.78\pm0.01$ \\
     2 & $0.58\pm0.04$ & $0.72\pm0.03$ & $0.77\pm0.03$ & $0.82\pm0.01$ \\
     3 & $0.64\pm0.05$ & $0.76\pm0.02$ & $0.81\pm0.02$ & $0.84\pm0.01$ \\
     4 & $0.65\pm0.04$ & $0.76\pm0.02$ & $0.79\pm0.03$ & $0.84\pm0.01$ \\
     5 & $0.66\pm0.03$ & $0.77\pm0.02$ & $0.82\pm0.02$ & $0.84\pm0.01$ \\
     \bottomrule
  \end{tabular}
  \caption{Comparison of Nearest-Neighbor performance for different memory bank sizes.}
  \label{table:abl_nn}
\end{table}

\begin{table}[tp]

  \centering
  \footnotesize
\begin{tabular}{cccccc}
    \toprule
    \makecell{$p$}     & $k=25\%$        & $k=50\%$      & $k=75\%$  & $k=100\%$\\ \midrule
     1 & $0.55\pm0.03$ & $0.57\pm0.02$ & $0.56\pm0.01$ & $0.52\pm0.04$\\
     2 & $0.60\pm0.04$ & $0.62\pm0.03$ & $0.58\pm0.03$ & $0.51\pm0.04$\\
     3 & $0.66\pm0.03$ & $0.70\pm0.02$ & $0.67\pm0.02$& $0.52\pm0.06$\\
     4 & $0.68\pm0.03$ & $0.68\pm0.03$ &  OOM & OOM\\
     5 & $0.68\pm0.02$ & OOM & OOM & OOM \\
     \bottomrule
  \end{tabular}
  
  \caption{Comparison of RPG for different $k$ and image pool sizes. 'OOM' denotes '\textbf{O}ut \textbf{O}f \textbf{M}emory'.}
  \label{table:radius}
\end{table}

\begin{table*}[t]
  \centering
  \begin{tabular}{llllllll}
  \toprule
  Methods&\multicolumn{1}{c}{Data}&\multicolumn{1}{c}{Right Lung}&\multicolumn{1}{c}{Left Lung}&\multicolumn{1}{c}{Heart}&\multicolumn{1}{c}{Right Clavicle}&\multicolumn{1}{c}{Left Clavicle}&\multicolumn{1}{c}{Mean}\\
    \midrule
    Baseline &\multirow{5}*{\rotatebox[origin=c]{90}{\makecell{JSRT\\$N_l=3$}}}&   $0.83 \pm 0.02$ & $0.81 \pm0.01$& $0.59\pm0.02$ &$0.47\pm 0.05$& $0.42 \pm 0.07$&$0.59\pm0.04$\\
    Pseudolabel$_{\tau = 0.8}$ && $0.86\pm0.02$ & $0.87\pm0.02$ & $0.65\pm0.10$ & $0.35\pm0.06$ & $0.28\pm0.06$ & $0.56\pm0.04$\\
    FixMatch$_{\tau = 0.8}$  &&\underline{$0.94\pm0.00$}&\underline{$0.93\pm0.00$}&\underline{$0.82\pm0.03$}&$0.50\pm0.09$&$0.44\pm0.14$&\underline{$0.71\pm0.05$}\\
    RPG       (Ours)                               &&$0.91\pm0.01$&$0.90\pm0.01$&$0.71\pm0.02$&\underline{$0.55\pm0.04$}&$\mathbf{0.55\pm0.03}$&\underline{$0.71\pm0.02$}\\
    RPG$^{+*}$ (Ours) &&$\mathbf{0.95\pm0.00}$&$\mathbf{0.95\pm0.00}$&$\mathbf{0.85\pm0.02}$&$\mathbf{0.60\pm0.09}$&\underline{$0.50 \pm 0.15$}& $\mathbf{0.77\pm0.05}$\\
    \midrule
    Baseline &\multirow{5}*{\rotatebox[origin=c]{90}{\makecell{Custom\\$N_l=2$}}}&
    \multicolumn{1}{c}{0.1105}&\multicolumn{1}{c}{0.0994}&\multicolumn{1}{c}{0.2335}&\multicolumn{1}{c}{\underline{0.0526}}&\multicolumn{1}{c}{\textbf{0.0256}}&\multicolumn{1}{c}{0.1043}\\
    Pseudolabel$_{\tau = 0.8}$  &&
    \multicolumn{1}{c}{0.2335}&\multicolumn{1}{c}{0.0847}&\multicolumn{1}{c}{0.0920}&\multicolumn{1}{c}{0.0000}&\multicolumn{1}{c}{0.0000}&\multicolumn{1}{c}{0.0820}\\
    FixMatch$_{\tau = 0.8}$  &&
    \multicolumn{1}{c}{0.0504}&\multicolumn{1}{c}{0.0463}&\multicolumn{1}{c}{0.0041}&\multicolumn{1}{c}{ 0.0000}&\multicolumn{1}{c}{0.0000}&\multicolumn{1}{c}{0.0246}\\
    RPG  (Ours)&&
    \multicolumn{1}{c}{\underline {0.6065}}&\multicolumn{1}{c}{\underline {0.4592}}&\multicolumn{1}{c}{\underline{0.5108}}&\multicolumn{1}{c}{0.0000}&\multicolumn{1}{c}{0.0000}&\multicolumn{1}{c}{\underline {0.3153}}\\
    RPG$^+$  (Ours)&&
    \multicolumn{1}{c}{\textbf{0.6326}}&\multicolumn{1}{c}{\textbf{0.4852}}&\multicolumn{1}{c}{\textbf{0.5636}}&\multicolumn{1}{c}{\textbf{0.0671}}&\multicolumn{1}{c}{\underline{0.0168}}&\multicolumn{1}{c}{\textbf{0.3531}}
    \\\bottomrule
\end{tabular}
  \caption{Performance comparison on JSRT and our extended annotations (Custom). $^*$ denotes a training of twice the iterations. Bold and underlined denote best and second best performance respectively.}
    \label{table:classes_jsrt}
\end{table*}

\begin{figure*}[b!]
    \centering
    \includegraphics[width=\linewidth]{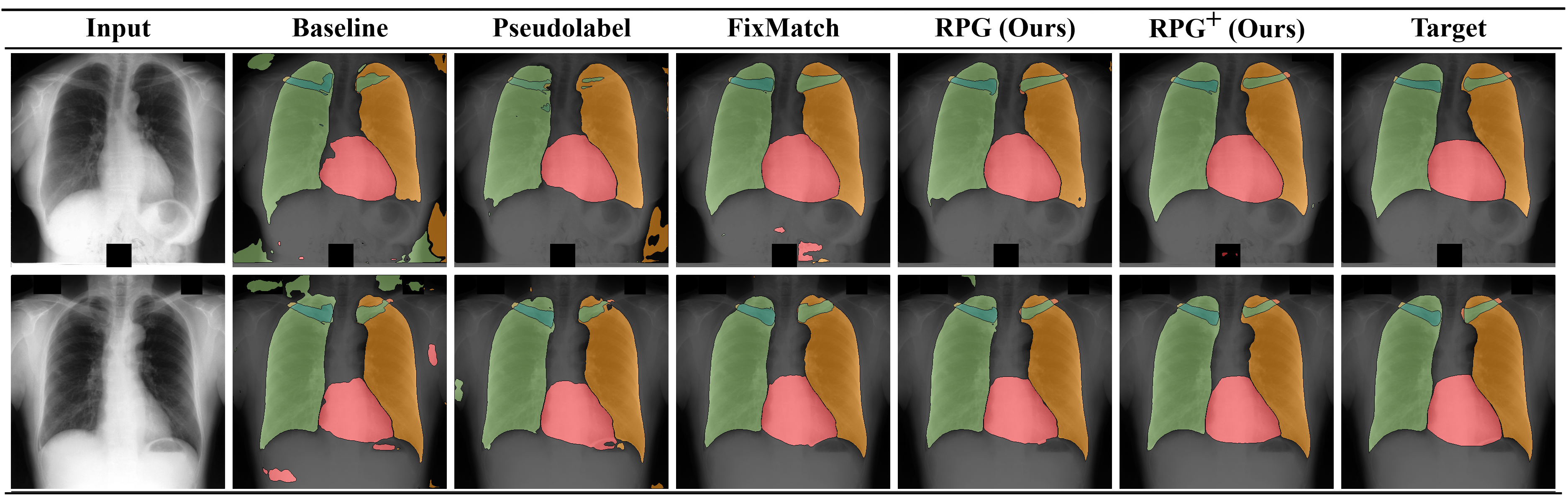}    
    \caption{Qualitative Segmentation Results on the JSRT~\cite{jsrt} dataset for $N_l=6$.}
   
     \label{fig:jsrt_images}
\end{figure*}

\noindent\textbf{Baselines and Methods.} First, we employ a standard UNet \emph{Baseline} using only the available annotated data as supervision. Due to the low amount of seen images, we run these models for the same amount of iterations instead of the same amount of epochs. Next, we take a look at the original \emph{Pseudolabel} method proposed by \cite{lee2013pseudo}, which we test for different thresholds $\tau$. Further, we compare a naive \emph{Nearest Neighbor} label assignment without a weighting function. We also compare against recent methods of \emph{MLDS}~\cite{reiss2021every}, which uses deep supervision paired with a Mean-Teacher~\cite{tarvainen2017mean} setup, and \emph{FixMatch}~\cite{sohn2020fixmatch} utilizing strongly and weakly augmented prediction comparisons.

\subsection{Ablation Studies}
\noindent\textbf{Pool Size for Pseudo-Label Assignment.}
To show the potential of nearest-neighbor-based pseudo-label generation for semantic segmentation, we investigate the segmentation performance for different combinations of pool size and the amount of annotated images. Table~\ref{table:abl_nn} shows that if only a single image is considered the segmentation ability of the network is rather poor but as we increase the number of images in the pool, we see a steady performance increase across all amounts of annotated images. We note that while increasing the pool size overall also positively affects performance, the improvement past $p=3$ is substantially less while the needed memory rises at a constant rate. Thus, we see $p=3$ as a good trade-off between performance and memory consumption and maintain it for all further experiments.

\noindent\textbf{Amount of Observed Neighbors.} 
We expand upon the nearest neighbor assignment in RPG with our weighting scheme. Therefore, we investigate the impact of the effective radius in feature space on our density-based class entropy weighting for different pool sizes $p$ and relative amounts of considered neighbors $k$ for $N_l=3$. We display the results in Table~\ref{table:radius} with each column representing the relative amount of all considered features in the reference pool $\mathcal{R}_\mathcal{P}$. Independently of the pool size, we see that increasing $k$ improves over just the nearest-neighbor assignments shown in the first column of Table~\ref{table:abl_nn}. The performance increases up until $50\%$ of all existing features where it peaks and falls off drastically for larger $k$'s. This indicates that an increased search radius is helpful but finding the optimal $k$ is difficult. Due to memory constraints we did not test larger $k$'s for bigger pool sizes.

\begin{table*}[t!]
  \centering
  \begin{tabular}{lllll}
  \toprule
  Methods&\multicolumn{1}{c}{$N_l=3$}&\multicolumn{1}{c}{$N_l=6$}&\multicolumn{1}{c}{$N_l=12$}&\multicolumn{1}{c}{$N_l=24$}\\
    \midrule
    Baseline &$0.15\pm0.07$&$0.27\pm0.08$&$0.35\pm0.06$&$0.49\pm0.05$\\
    \midrule
    IIC~\cite{ji2019invariant}                  &\underline{$0.22\pm0.09$}&$0.32\pm0.07$&$0.41\pm0.07$&$0.53\pm0.06$\\
    Perone and Cohen-Adad (2018) &$0.21\pm0.09$&$0.31\pm0.10$&$0.39\pm0.07$&$0.50\pm0.08$\\
    MLDS~\cite{reiss2021every}                   &$0.16\pm0.15$&\underline{$0.35\pm0.11$}&\underline{$0.54\pm0.09$}&$\mathbf{0.59\pm0.07}$\\
    RPG (Ours)                                &$0.21\pm0.10$&$0.30\pm0.08$&$0.45\pm0.08$&\underline{$0.54\pm0.08$}\\
    RPG$^+$ (Ours)                       &$\mathbf{0.31\pm0.11}$&$\mathbf{0.45\pm0.10}$&$\mathbf{0.55\pm0.08}$&$\mathbf{0.59\pm0.08}$\\\midrule
    Full Access ($N_l=415$) &\multicolumn{4}{c}{$0.62\pm0.05$}\\\bottomrule
    
  \end{tabular}
  \caption{Performance comparison on Retouch. Bold and underlined denote best and second best performance respectively.}
    \label{table:retouch}

\end{table*}

\subsection{Quantitative Results}
\begin{figure}[t]
    \centering
    
    \includegraphics[width=\linewidth]{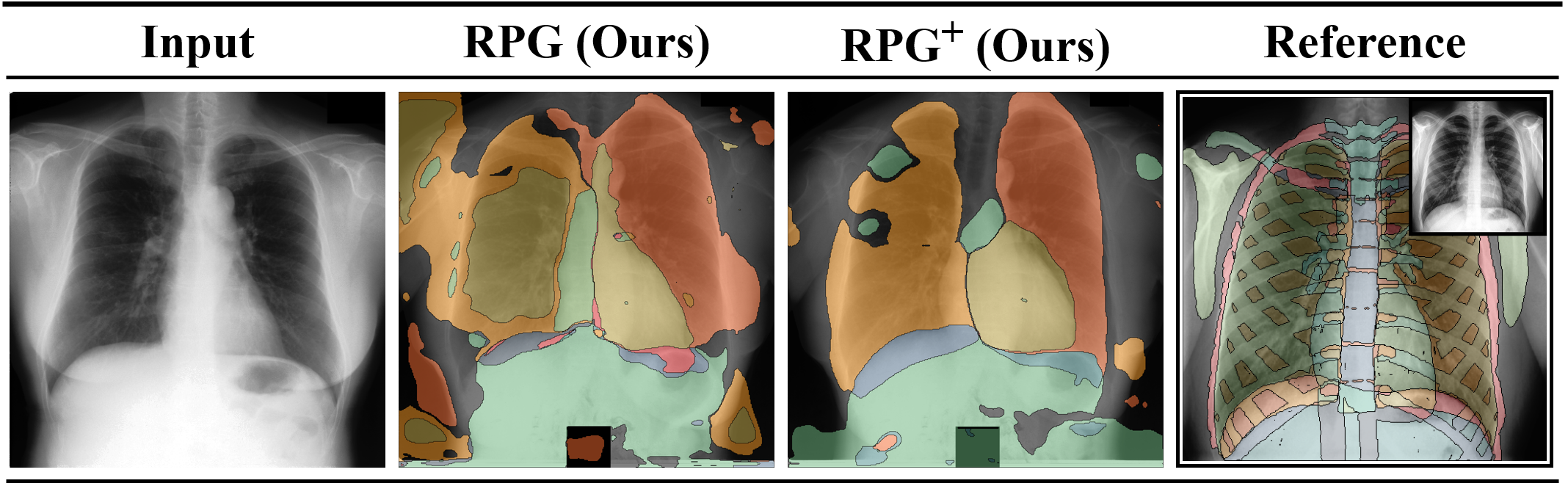}
    \caption{Qualitative Segmentation Results on extended anatomical x-ray annotations. }
    \label{fig:jsrt_extended_images}
\end{figure}

\begin{figure}[bp!]
    \centering
    \includegraphics[width=\linewidth]{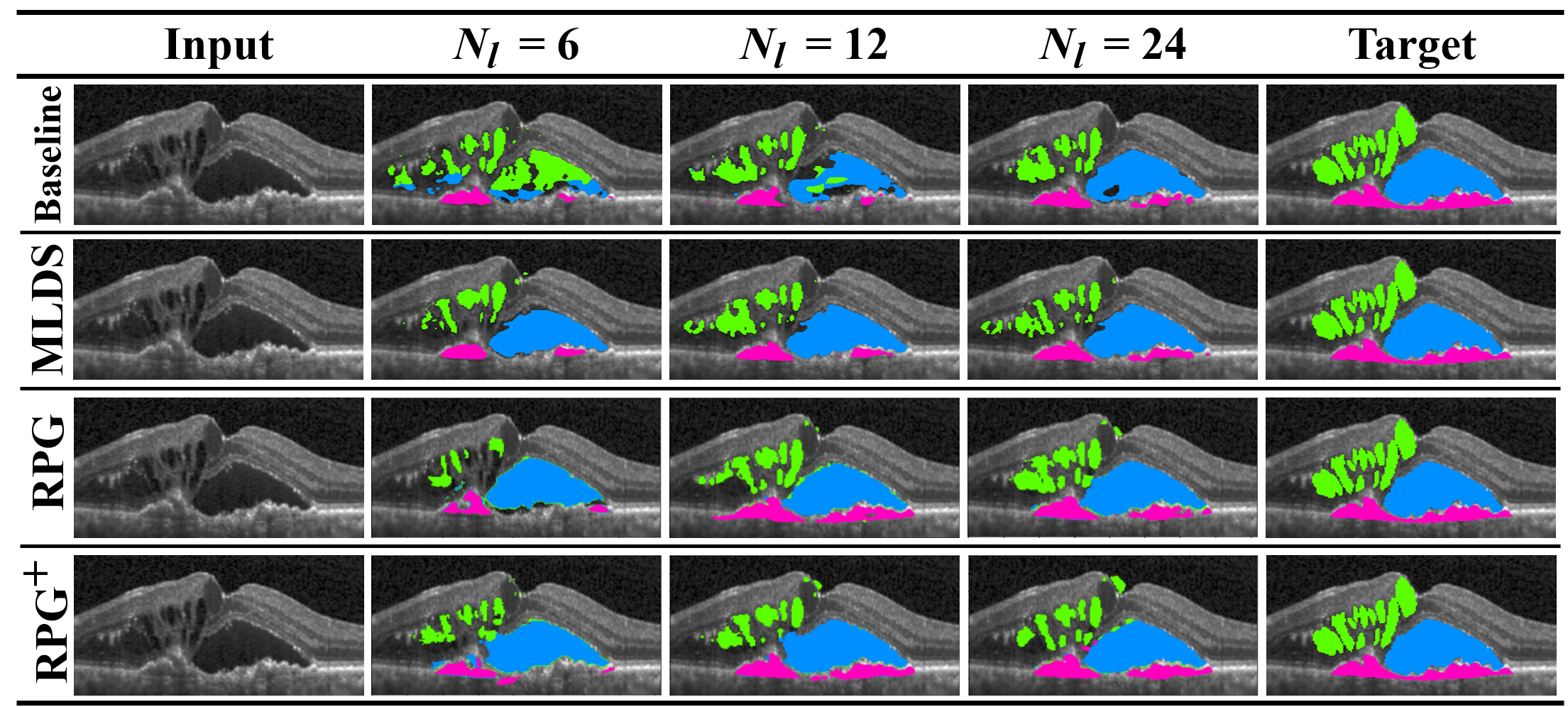}
    \caption{Qualitative Results on RETOUCH.}
    \label{fig:retouch}
\end{figure}

\noindent\textbf{Results on JSRT.} 
In Table~\ref{table:jsrt_results}, we display the mIoU of various methods for multi-label anatomy segmentation. 
Standard pseudo-labeling shows varying performance depending on the chosen threshold, and while it performs well for more annotated images,  the performance even falls below the baseline in the low data case for any chosen threshold. Nearest Neighbor pseudo-labels improve over the baseline for few samples by $3$-$5\%$ but show slightly worse results for 24 annotated examples. FixMatch gains $12\%$ above the baseline for $N_l=3$, but it struggles for more annotations despite taking longer to converge. Both FixMatch and standard Pseudolabeling show varying performance for different $\tau$. Our proposed $RPG$ performs equally to FixMatch for smaller $N_l$ and outperforms it for larger $N_l$s while not using strong augmentations. We further see that the integration of strongly augmented images in $RPG^+$ improves the performance of $RPG$ for all settings gaining $18\%$ over the baseline for $N_l=3$ and matches fully supervised performance with six labeled samples.

We further demonstrate the class-wise performance for three annotated samples on the top of Table~\ref{table:classes_jsrt}. We see that the baseline as well as pseudo labeling struggle with less common classes like the clavicles. FixMatch shows considerable improvements for the classes with more annotated pixels, while the performance for the clavicles only slightly improves. $RPG$ also improves over the baseline for heart and lungs,
but shows significant improvements for the difficult clavicles. Furthermore, $RPG^+$ combines the aspects of $RPG$ and augmentation-based consistency regularization, which noticeably improves all categories apart from the \textit{right clavicle} with gains up to absolute $26\%$ over the baseline. 
We display segmentation predictions in Fig.~\ref{fig:jsrt_images}, where class-wise shortcomings of the different methods become visible.

\noindent\textbf{Results on Extended JSRT.}
In the bottom half of Table~\ref{table:classes_jsrt}, we display the results when using our fine-grained annotations of JSRT. The baseline of training simply on the annotated images achieves $10.43\%$ mIoU. Both prediction-based pseudolabeling methods struggle in this complex low data environment and perform worse than the baseline. In contrast, RPG manages to correctly predict classes of the super-categories \textit{Left Lung, Right Lung} and \textit{Heart} leading to a mIoU of $31.53\%$ thus improving upon the baseline by absolute $21.10\%$. RPG$^+$ slightly boosts this further to a mIoU of $35.31\%$.
We display the extended anatomy segmentations in Fig.~\ref{fig:jsrt_extended_images}. We see RPG$^+$  managing to reconstruct the lung-subcategories, ventricles of the heart and the sub-diaphragm, but struggling with explicit predictions of bone structures.

\noindent\textbf{Results on Spectralis.}
We display our results for the Spectralis dataset in Table~\ref{table:retouch}. Here, we also see $RPG$ outperforming the baseline for all considered $N_l$, thus showing its usability for the multi-class segmentation setting. $RPG^+$ noticeably outperforms other methods especially for the low data schemes of $N_l=3$ and $N_l=6$ by up to 15\% while having the same performance as MLDS for $N_l=24$. We display qualitative comparisons in Fig.~\ref{fig:retouch}.

\section{Conclusion}
In this work, we proposed a novel way of generating supervision for segmentation. We use labeled images as reference material, match pixels in an unlabeled image to their semantic counterparts, and allocate the corresponding label seen in the reference. This way, we do not fall into pitfalls common with prediction-based pseudo-labeling such as confirmation bias.  Since no additional networks or alterations to a given architecture are necessary, our proposed method can easily be plugged into any existing framework. We argue that this way of pseudo-label generation is especially fitting for medical image analysis due structural similarity provided by underlying anatomical structures. We demonstrate the effectiveness of our approach through extensive experiments on chest X-ray anatomy segmentation and retinal fluid segmentation. We achieve fully supervised performance with only a handful of samples, thus, cutting the annotation cost by 95\%. Our code and additional information are available in the supplementary.

\section{Acknowledgements}
The present contribution is supported by the Helmholtz Association under the joint research school “HIDSS4Health – Helmholtz Information and Data Science School for Health”. We further thank Vincent Braun for medical feedback and R\'emi Delaby for fruitful theoretical discussions.

\bibliography{aaai22}

\end{document}